\documentclass[namedreferences]{kluwer}
\usepackage{graphicx}

\newcommand{\ecsa}{\mbox{${\mathrm{erg\,cm^{-2}\,s^{-1}\,\AA^{-1}}}$}}
\newcommand{\La}{\mbox{${\rm Ly\alpha}$}}
\newcommand{\Msun}{\mbox{$M_{\odot}$}}
\newcommand{\kms}{\mbox{$\mathrm{km\,s^{-1}}$}}
\newcommand{\Msec}{\mbox{$M_{\mathrm{sec}}$}}
\newcommand{\Mwd}{\mbox{$M_{\mathrm{wd}}$}}
\newcommand{\Twd}{\mbox{$T_{\mathrm{wd}}$}}

\newcommand{\Ion}[2]{#1{\,\scriptsize #2}}
\newcommand{\msy}{\mbox{$\mathrm{\Msun\,yr^{-1}}$}}

\begin{document}
\begin{article}
\begin{opening}
\title{Ultraviolet studies of interacting binaries}

\author{Boris T. \surname{G\"ansicke$^1$}, Domitilla \surname{de
Martino$^2$ }, Thomas R. \surname{Marsh$^1$}, Carole
A. \surname{Haswell$^3$}, Christian \surname{Knigge$^4$}, Knox
S. \surname{Long$^5$}, Steven N. \surname{Shore$^6$}}

\runningauthor{G\"ansicke et al.}
\runningtitle{Interacting Binaries in the UV}

\institute{
$^1$ Department of Physics, University of Warwick, Coventry CV4 7AL,
  UK\\
$^2$ INAF - Osservatorio di Capodimonte, Via Moiariello 16, 80131 Napoli, 
Italy\\
$^3$ Department of Physics and Astronomy, The Open University, Milton
Keynes MK7 6AA, UK\\
$^4$ School of Physics and Astronomy, University of Southampton,
Southampton SO17 1BJ, UK \\
$^5$Space Telescope Science Institute, 3700 San Martin Drive,
Baltimore, MD 21218\\
$^6$ Dipartimento di Fisica, Universit\'a di Pisa, Largo Pontecorvo 2,
56127 Pisa, Italy }


\begin{abstract}
Interacting Binaries consist of a variety of stellar objects in
different stages of evolution and those containing accreting compact
objects still represent a major challenge to our understanding of not
only close binary evolution but also of the chemical evolution of the
Galaxy. These end-points of binary star evolution are ideal
laboratories for the study of accretion and outflow processes, and
provide insight on matter under extreme physical conditions. One of
the key-questions of fundamental relevance is the nature of SN\,Ia
progenitors. The study of accreting compact binary systems relies on
observations over the entire electromagnetic spectrum and we outline
here those unresolved questions for which access to the ultraviolet
range is vital, as they cannot be addressed by observations in any
other spectral region. 
\end{abstract}

\end{opening}

\section{Scientific background and astrophysical context}
The 20th century saw an impressive leap in the theory of stellar
evolution~--~leading from not even knowing what source of energy
powers the Sun to the extremely detailed models of stellar structure
and evolution available today. A number of the present-day key
research areas, e.g. galaxy evolution, are deeply rooted in our
understanding of stellar evolution.  However, while we may feel
comfortable about our understanding of single stars, observational
evidence collected throughout the last few decades makes it
increasingly clear that the majority of all stars in the sky are born
in binaries, of which many will interact at some point in their lives
\cite{iben91-1}.  Virtually all of the most exotic objects in the
Galaxy are descended from such binary stars, including binary pulsars,
all the galactic black-hole candidates, low-mass X-ray binaries
(LMXB), millisecond pulsars, cataclysmic variables (CVs), symbiotic
stars, and many others.  Binary stars are important in many other
contexts, too.  Sub-dwarf\,B stars, which now appear to be another
product of binary evolution, dominate the ultraviolet light of old
galaxies. The Type Ia supernovae, among the most important `standard
candles' in the determination of extragalactic distances on a
cosmological scale, are thought to arise from exploding white dwarfs
driven over their Chandrasekhar mass limit by accretion from a
companion star. Even the class of short gamma-ray bursts, the most
powerful explosions in the Universe, may be related to the merging of
two neutron stars, again products of binary star evolution.

Interacting binary stars are showcases of the processes of mass
accretion and outflow, exhibiting a variety of phenomena such as
accretion discs, winds, collimated jets and magnetically controlled
accretion flows, thermal disc instabilities, and both stable and explosive
thermonuclear shell burning. The plasma conditions in these accretion
structures span a huge range of physical conditions, including
relativistic environments and extreme magnetic field
strengths. Consequently, interacting binaries are also extremely
versatile plasma physic laboratories.

Despite their great importance for a vast range of astrophysical
questions, our understanding of close binary stars and their evolution
is still very fragmentary. The ultraviolet (UV) is of outmost
importance in the study of interacting binaries, as a large part of
their luminosity is radiated away in this wavelength range, and, more
importantly, as the UV hosts a multitude of low and high excitation
lines of a large variety of chemical species. These transitions can be
used both as probes of the plasma conditions, as well as tracers of
individual components within the binaries through time-resolved
spectroscopy. Moreover, the physical status of the binary components and
in particular the accreting white dwarf primaries in cataclysmic
variables, symbiotic stars, and double-degenerate binaries can be
easily isolated and studied in the UV range. Even though 
substantial scientific progress has been achieved throughout the last
three decades, primarily using the \textit{International Ultraviolet
Explorer} (\textit{IUE}), the \textit{Hubble Space Telescope}
(\textit{HST}), and the \textit{Far Ultraviolet Spectroscopic Explorer}
(\textit{FUSE}), these are still the early days of UV astronomy of
interacting binaries, and many key questions are yet without answer. 
Here we outline the enormous potential that a major UV observatory has
for our understanding of interacting binaries, and how the expected
findings related to much wider astrophysical contexts, including
galaxy evolution and cosmology.

\section{Accreting white dwarfs}

\subsection{The complex interplay between stellar properties and binary evolution} 

Compared to their isolated relatives, the evolution of white dwarfs in
interacting binaries is much more complex, and closely related to the
evolution of the binary as a whole, and hence understanding close
binary stellar evolution is impossible without detailed knowledge of
the properties of the white dwarf components in these stars.  The most
abundant type of mass-transferring binaries containing a white dwarf
are the cataclysmic variables (CVs), which have mass transfer rates in
the range $10^{-11}-10^{-9}$\,\msy. The accretion of this material and
its associated angular momentum affects practically all fundamental
properties of CV white dwarfs.

Compressional heating is depositing energy in the envelope and the
core of the white dwarf, effectively compensating the secular cooling,
with the result that accreting white dwarfs are substantially hotter
than isolated white dwarfs of comparable age and
mass. \inlinecite{townsley+bildsten02-2,townsley+bildsten03-1} have
shown that the white dwarf temperature can indeed be used to establish
a measure of the long term average of the accretion rate sustained by
the white dwarf. As the secular average accretion rate is directly
related to the rate at which the binary is losing orbital angular
momentum measuring this parameter is of fundamental importance for any
theory of close binary evolution.

Accretion will increase the mass of the white dwarf. Eventually, if
nothing else happens and the mass supply of the companion star is
sufficient, this will drive the white dwarf over its Chandrasekhar
mass limit, and it may turn in into a supernova Type Ia. Accreting white
dwarfs as possible SN\,Ia progenitors are discussed in more detail in
Sect.\,\ref{s-snia} below. However, in most CVs the accreted hydrogen
layer will thermonuclearly ignite once the density and temperature
exceed the critical condition. This hydrogen shell burning is typically
explosive, observationally designated as a classical nova, and ejects
a shell of material into space (see Sect.\,\ref{s-novae}). As the
critical mass of the accreted hydrogen-rich layer is fairly low
($\sim10^{-5}-10^{-3}$\,\Msun), a CV will undergo hundreds to
thousands of nova explosions. Currently, it is not clear what the mass
balance during the nova event is, i.e. whether the amount of ejected
material is equal to or even exceeds the mass of accreted material,
and, hence, the long-term evolution of the white dwarf mass is not
known.

Chemical abundances of white dwarf surfaces can be affected by
accretion and greatly modified by nova explosions. While a roughly
solar composition is expected for a freshly accreted white dwarf
atmosphere, many CVs were recently found to possess an unexpected wide
variety of departures from (solar) abundances \cite{sion99-1}, opening
new horizons in the current understanding of binary evolution. Indeed
while in single white dwarfs metallic species and their abundance
reveal processes which oppose diffusion, those in cataclysmic
variables show a mix of chemical species and abundances that cannot
result from accretion from a normal secondary star, thus pointing
towards a thermonuclear activity in their past evolution. The
hypothesis of CNO processing as the source of the abundances has been
further supported by the detection of proton-capture material by
\inlinecite{sionetal97-1}.  This has great implications for CV
evolution and contributions to the heavy element content of the
interstellar medium (see also Sect.\,\ref{s-snia}).

Rotation rates of non-magnetic white dwarfs in CVs were unknown prior
the \textit{HST} era and its advent opened a new topic in close binary
evolution. Global rotational velocities are now measured for a handful
of dwarf novae systems \cite{sion99-1} and were found to be much
larger (300--1200\,\kms) than the few tens of \kms\ in isolated white
dwarfs, implying that accretion efficiently spins-up the primaries.
However the measured rates are much lower than expected on the basis
of the amount of angular momentum accreted during their characteristic
lifetimes \cite{livio+pringle98-1,kingetal91-1} suggesting that part
of the accreted angular momentum is removed from the white dwarf
during the expanded envelope mass loss phase which follows a nova
eruption. This independently suggests that also dwarf nova experience
nova outbursts and then return to be dwarf novae again, as suggested
by the cyclic evolution scenario. Although this result has an enormous
evolutionary implication, it is based on only 5 CV white dwarfs for
which reliable rotation rates could be determined.

\begin{figure}
\centerline{\includegraphics[angle=270,width=0.95\textwidth]{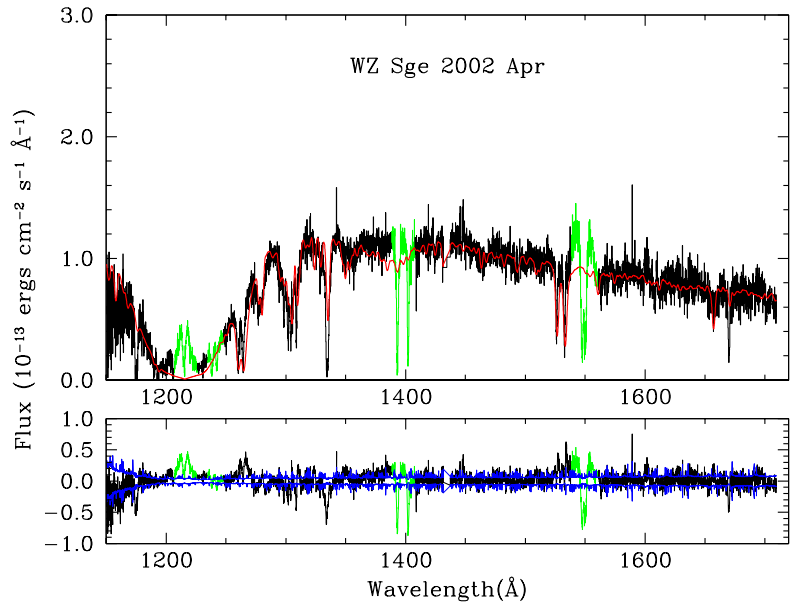}}
\caption{High-quality UV spectroscopy of accreting white dwarfs in CVs
  is necessary to determine their temperature, mass, rotation rate,
  and atmospheric abundances from detailed model atmosphere fits. Only
  a very limited number of CVs has been bright enough to be studied
  with the STIS high resolution grating, such as e.g. WZ\,Sge
  (from \protect\opencite{longetal04-1}).}
\label{f-cvwd}
\end{figure}

Whereas accretion alters the properties of white dwarfs in interacting
binaries, some of the white dwarf characteristics will in turn deeply
affect the accretion process~---~e.g. the mass of the white dwarf
defines the depth of the potential well, and, thereby, the amount of
energy released per accreted gram of matter, the rotation rate of the
white dwarf determines the luminosity of the boundary layer, i.e. the
interface between the inner accretion disc rotating at Keplerian
velocities and the white dwarf itself, and finally the magnetic field
of the white dwarf determines the accretion geometry.

The observational study of accreting white dwarfs can only be carried
out in the UV, as the emission from the accretion flow
dilutes or even completely outshines the white dwarf at optical
wavelengths. Because of the faintness of most CV white dwarfs, the
number of systems for which medium-resolution ($\simeq1-2$\,\AA)
spectroscopic data, adequate for temperature measurements, has been
obtained is $\simeq35$ (\opencite{sion99-1},
\opencite{szkodyetal02-1}, \opencite{araujo-betancoretal05-2})~--~out
of a total of $\sim1000$ CVs known
\cite{downesetal01-1}. High-resolution ($\simeq0.1$\,\AA) UV
spectroscopy necessary for accurate abundance and rotation rate
determinations has been obtained only for a handful of systems, most
noticeably for the nearest CV, WZ\,Sge (Fig.\,\ref{f-cvwd}, see
e.g. \opencite{sionetal01-1}, \opencite{sionetal03-2},
\opencite{longetal04-1}, \opencite{welshetal03-1}), at the expense of
$>20$ \textit{HST} orbits.

\medskip
\textbf{Future prospectives of UV astronomy.}  In order to fully
assess the interrelation between the white dwarf properties and the
evolutionary state of CVs, a sufficiently large number of systems has
to be observed. Mapping out the parameter space (\Twd, \Mwd,
abundances and rotation rate of the white dwarf, as well as the binary
orbital period) will eventually require adequate data for 100--200
systems. Temperature measurements need a broad UV wavelength coverage,
optimally from the Lyman edge down to ~3000\,\AA, at a low resolution
($R\simeq1000-2000$). Abundance/rotation rate measurements rely on
medium-resolution ($R\simeq20000$) spectroscopy covering a sufficient
number of transitions; the traditional range 1150\,\AA--1900\,\AA\ is
adequate even though similar capabilities below $\La$ would be
desirable. Throughput is the crucial need for this science, as typical
flux levels are a few  $\simeq10^{-16}$\,\ecsa.

\subsection{\label{s-snia}Accreting white dwarfs as likely SN\,Ia progenitors}
Whereas SN\,Ia are routinely used as beacons at cosmological distances
\cite{filipenko04-1}, and generally associated with the thermonuclear
disruption of a carbon-oxygen white dwarf \cite{livio01-1}, the nature
of their progenitors remains elusive.  Two different channels of
SN\,Ia progenitors are currently most favoured
\cite{yungelson+livio00-1}.  In the double-degenerate channel two
white dwarfs spiral in under the effect of gravitational radiation
until they finally merge, exceeding the Chandrasekhar mass
limit. Intensive optical surveys have been carried out for this type
of SN\,Ia progenitors, most recently by \cite{napiwotzkietal01-1},
identifying a few potential SN\,Ia progenitor candidates. In the
single-degenerate channel a white dwarf accretes from a main-sequence
companion. However, as outlined above, most white dwarfs accreting
hydrogen-rich material will go through classical nova explosions and
grow only little (or even shrink) in mass. Only if the white dwarf is
accreting at a rate sufficiently high to sustain {\it steady-state
hydrogen shell burning}~--~the accreted hydrogen is thermonuclearly
processed at the rate it is accreted. These objects have been
predicted (\opencite{sharaetal77-1}, \opencite{iben82-1},
\opencite{fujimoto82-1}) and first found in the EINSTEIN X-ray survey
of the Magellanic clouds \cite{longetal81-1}, even though it took a
fair amount of time to identify their true nature
\cite{vandenheuveletal92-1}.  Based on their observational
hallmark~--~a very large luminosity in soft X-rays, these objects are
coined supersoft sources, or more appropriately supersoft X-ray
binaries \cite{gaensickeetal00-3}.  The high accretion rates that are
necessary to fuel the steady-state shell burning in supersoft X-ray
binaries can be provided by a Roche-lobe filling main sequence star if
its mass is similar to or exceeds that of the white dwarf. As mass is
transfered from the more massive to the less massive star, the binary
period shrinks as a consequence of angular momentum conservation,
stabilising or even enhancing the mass loss of the donor star. The
mass transfer ensues on a time scale which is too short for the donor
star to adjust its thermal structure, and in an evolutionary jargon
supersoft X-ray binaries are known as thermal time scale mass transfer
(TTSMT) CVs.  In the absence of nova eruptions, the white dwarfs in
TTSMT CVs grow in mass, and will, if the donor star provides a
sufficient amount of material, surpass the Chandrasekhar limit and
potentially explode in a SN\,Ia (\opencite{distefano96-1},
\opencite{starrfieldetal04-1}).

If the donor star in a TTSMT runs out of fuel before the white dwarf
reaches the Chandrasekhar mass limit, the mass ratio will eventually
flip with the donor star being less massive than the white
dwarf. Consequently, the mass transfer rate decreases and the shell
burning ceases. From this point on, the system will evolve and look
(at a first glance) like a normal CV~--~with the dramatic difference
that normal CVs contain main-sequence donor stars, whereas post-TTSMT
CVs contain the CNO processed core of the previously more massive
star. \inlinecite{schenkeretal02-1} suggest that a significant
fraction (up to 1/3) of all present-day CVs may actually have started
out with a companion more massive than the white dwarf, and underwent
a phase of TTSMT. A recent \textit{HST}/STIS snapshot survey of 70 CVs showed
that $\sim10$\,\% of the systems display a significantly enhanced N/C
abundance ratio, which suggests that these systems went through a
phase of TTSMT \cite{gaensickeetal03-1}.  So far, not a single
\textit{progenitor} of supersoft X-ray binaries/TTSMT CVs has been identified.

\medskip
\textbf{Future prospectives of UV astronomy.}  Mapping out the
population of failed SN\,Ia\,=\,post-TTSMT CVs will require low
($R\simeq1000-2000$) resolution spectroscopy of several 100 CVs down
to $\simeq 10^{-16}$\,\ecsa .  Based this large sample, it will be
possible to determine the orbital period distribution of post-TTSMT
CVs with respect to the ``normal'' systems, model their evolution, and
finally extrapolate these population models to the regime of true
SN\,Ia progenitors. Follow-up the brighter ones at high resolution
($R\simeq20000$) will be necessary in order to determine their
detailed properties. This will also help to answer the very important
question on whether the white dwarfs in these systems have grown in
mass, i.e. are they more massive than in normal CVs?

Equally important is the search for true SN\,Ia progenitors. However,
as the TTSMT phase is very short the chance of finding systems in this
stage is small~--~in fact, in our Galaxy (where absorption in the
plane further decreases the probability of finding such systems) only
two supersoft X-ray binaries are known. Supersoft X-ray binaries can
be located in local group galaxies using high spatial resolution
($\simeq1"$) X-ray missions such as Chandra~--~however, X-ray
data alone is typically insufficient to determine the properties of
the objects. UV observations can substantially defeat crowding
problems (see Sect.\,\ref{s-globulars}), and are well-suited to obtain
fundamental parameters such as orbital periods. This implies large
aperture high spatial resolution UV imaging capabilities (supersoft
X-ray binaries in M31 have $V\simeq23$).

A so far entirely unexplored potential is the search for
\textit{SN\,Ia pre-progenitors}, i.e. detached white dwarf/main
sequence binaries with $\Msec>1.6\,\Msun$ (\opencite{langeretal00-1},
\opencite{han+podsiadlowski04-1}). In the optical, these systems will
be entirely dominated by the main-sequence star, and follow-up
UV studies of main-sequence stars with UV excess (identified
e.g. in the GALEX survey) will be necessary to identify them.

\subsection{\label{s-novae} The nova phenomenon}

Novae are the most spectacular phenomenon encountered in CVs and
represent key objects to understand a wide variety of physical
conditions of accreting matter including super-Eddington regimes and
the interaction of ejecta in the interstellar medium and its chemical
evolution.  They are fundamental standard candles up to the Local
Group, having hence important implications for cosmological distance
calibrations.

Despite the enormous observational effort of the past 15 years,
especially with multi-wavelength campaigns and datasets, there remain
two fundamental uncertainties: what are the masses and structures of
the ejecta and what drives the mass loss during the outburst?
Radiative processes, which might be the source of a stellar wind
during the ejection phase (e.g. \opencite{hauschildtetal94-1}) depends
on the abundances (and therefore the details of the spectral evolution
during the initial stages) and the bolometric luminosity.  Explosions
are powered directly by decays of radioactive isotopes generated
during the thermonuclear runaway following the initial envelope
expansion, but any subsequent mass loss must be driven by the match of
the flux distribution with the envelope opacities \cite{shore02-1}.
The UV, now inaccessible to observation, is the driving spectral
region for the phenomenological analysis of novae.  Only in this
region it is possible to directly probe the properties of the ejecta -
abundances, structure, mass~--~and determine the energetics of the
thermonuclear runaway.  The reasons are simply that the photometric
behaviour is driven at all wavelengths longer than the UV by
bolometric flux redistribution from the evolving central remnant white
dwarf and that in the UV we can measure the resonance transitions off
the dominant ions throughout the first few months of outburst.  To
date, only novae in the Galaxy and the LMC have been observed.

Novae have been used as distance calibrators for nearly a century
through the maximum magnitude - rate of decline (MMRD) relation, but
the origin of this relation has only recently been understood.  As the
ejecta expand, the rapid and enormous increase in the opacity from
recombination-driven strengthening of the line absorption
redistributes flux into the optical.  But the correspondence between
these two regions, the completeness of the redistribution process,
depends on the details of the ejecta \textit{filling factors}.  If the
ejecta initially fragment and/or if they are not spherical in the
earliest stages, the process will be less efficient and the observed
maximum at longer wavelengths will be altered.  Without the UV, it is
impossible to determine the bolometric luminosity and therefore to
constrain its constancy.

The two principal classes of \textit{classical} novae are
distinguished by their abundances, which reflect the composition
differences of the accreting white dwarf (CO and ONe).  The most
extreme explosions may produce significantly altered abundance
patterns and there is an indication that the ejecta for both of these
types are also helium enriched.  Without the UV to provide access to
resonance lines for the relevant ions, abundance studies~--~and the
determination of structure~--~are limited by uncertainties in the
equation of state for the ejecta.

Among recurrent novae, the two classes; those in compact, 
cataclysmic-like systems and those with red giant 
companions, appear to have very low mass ejecta (in 
agreement with current models) but with abundance patterns 
that suggest helium enrichment.  The UV is the only way to 
study the ejecta in the optically thick phases (which last 
only a matter of days) to obtain unambiguously the 
abundances.  It also is not clear whether these systems 
show discs, or winds, during quiescence.  Finally, the 
interaction between the expanding ejecta and the winds in 
the symbiotic-like systems (with red giant companions) can 
only be studied effectively at high spectral resolution in the UV 
where the resonance lines and continuum of the white dwarf 
are accessible.

\medskip
\textbf{Future prospectives of UV astronomy.} With increased aperture,
especially in the 4--6 meter range, and high spectral resolution
(10\,000 or higher), it would be possible to study novae throughout the
Local Group, especially M\,31 in which the full range of novae appear
to occur.  Novae are important contributors to several rare isotopes,
especially $^{22}$Ne and possibly $^{22}$Na, and also may be important
in ionising galactic halos (thus being important for understanding the
halo ionisation and properties of Ly$\alpha$ Forest systems formed therein).
Since they are recurrent phenomena, on many timescales, and remain hot
for long periods they may be important for understanding the UV upturn
in elliptical galaxies (they can mimic post-AGB stars, for example).
Finally, as bright, transient UV sources, they provide probes of the
interstellar medium throughout their host galaxy. Also, the transition
from the super-soft phase into the UV is essential but it has always
been extremely difficult to determine.  Furthermore the determination
of chemical abundances in the ejecta is a fundamental parameter to
test theories on the processes that lead to the nova phenomenon and to
understand the state of the binary system. Indeed in no system these
abundances were found to be solar-like
(\opencite{selvelli+gilmozzi99-1}) having important implications in
the chemical evolution of interstellar medium.

\subsection{\label{s-symbiotic} Symbiotic stars}

The symbiotic systems are exotic and intriguing interacting
binaries. The nature of the accreting hot companion was longly debated
and proofs that the hot accreting companion is most likely a white
dwarf and not a main sequence star were provided by UV observations
(e.g. \opencite{erikssonetal04-1}). These systems however differ from
the CVs because of their wider orbits, with orbital periods from a few
to a few dozen years, and because the white dwarf accretes from
the stellar wind of a late-type giant rather than through Roche lobe
overflow from a main sequence star. The wind from the cool star is
ionised by the radiation from the white dwarf resulting in the
characteristic combination of sharp nebular emission lines and
molecular absorption bands in their UV and optical spectra
\cite{birrieletal00-1}. An increasing number of symbiotic stars are
also found to show nova outbursts. In these systems, despite the much
smaller outburst amplitudes compared to those observed in novae, the total
energy associated with the outburst may significantly exceed that of a
classical nova. Currently very little is known about the line-emitting
regions associated with the outburst of a symbiotic nova because of
the long timescales to reach the maximum and the very much slower
decays \cite{rudyetal99-1}.  Symbiotics also fall in the category of
the supersoft X-ray sources \cite{greiner96-1} making them potential
SN\,Ia progenitors \cite{hachisuetal99-1}.  Furthermore only a handful
of symbiotics have been monitored trough their outbursts in the UV,
where the evolution of the hot accreting object can be best followed
and from which the energetics of the process can be best studied and
linked to the soft X-ray emission \cite{gonzalez-riestraetal99-1}.

Among the different classes of interacting binaries discussed
in this paper, symbiotic stars are by far the physically largest
objects, and future UV/optical interferometric missions with
sub-milliarcsecond spatial resolutions will be able to resolve the
two stellar components, as well as the wind / accretion flow from
the companion star, and possibly an accretion disc around the
compact star. Carrying out such studies in the different UV
resonance lines will allow a detailed mapping of the ionization
structure in the accretion flow.

\medskip
\textbf{Future prospectives of UV astronomy.}  To identify the hot
accreting component and to determine the UV luminosity and its
evolution during outbursts the construction of SED over a wide
spectral range is necessary. This requires low dispersion ($\sim2000$)
spectroscopy in the desirable range from the Lyman limit down to
3400\,\AA. Our knowledge of the population of symbiotic novae in our
Galaxy and Local group will enormously improve with UV imaging
capabilities as described in Sect.\,\ref{s-snia} and
\ref{s-globulars}. Furthermore the study of emission lines mapping a
wide variety of physical conditions of the accreted matter and outflow
need moderate-to-high dispersion spectroscopy ($R\sim10\,000-20\,000$)
in the FUV range. UV imaging at at sub-milliarcsec resolution is
required to physically resolve the stellar components and accretion flow.

\subsection{Accretion flows in magnetic systems}

Accretion can be greatly influenced by the presence of magnetic fields
of the primary star. In those CVs where the white dwarf is strongly
magnetised ($B>10^5 - 10^8$\,G) important modifications of the
accretion flow occur already at the distance of the donor star. The
formation of an accretion disc is prevented in the high field systems
($B>10$\,MG) or truncation of the accretion disc can occur in
moderately ($B<5$\,MG) magnetised CVs. Hence, the wide range of
accretion patterns encountered in mCVs allows to test different
physical conditions of accretion flow and X-ray irradiation. In
particular, previous observations of magnetic systems have shown that
the FUV continuum is dominated by the X-ray irradiated white dwarf
pole \cite{gaensickeetal95-1}, while the accretion funnel down to the
post-shock regions contributes in the NUV continuum and in the FUV
emission lines of CNO. The truncated disc is also a source of UV
continuum (\opencite{haswelletal97-1}, \opencite{demartinoetal99-1},
\opencite{eisenbartetal02-1}, \opencite{belleetal03-1}).  However
there are still important open questions on the physical conditions
(kinematics, temperature and density) of the accretion flow:

(1) A strong potential to diagnose ionised gas is provided by the
resonance FUV emission lines (CNO), as their FWZI $\sim
2000-3000$\,\kms\ clearly indicates that they map the accretion flow
down to the white dwarf surface.  However, while past (low resolution)
UV observations have allowed significant progress in the understanding of
emission line formation ruling out collisional ionisation and strongly
favouring photoionisation models \cite{maucheetal97-1}, there is still
a great uncertainty in theory, as they cannot simultaneously account
for all the line flux ratios observed in CVs. Among the magnetic
systems there seems to be a higher ionisation efficiency in the hard
X-ray intermediate polar systems with respect to the soft X-ray polar
systems \cite{demartino99-1} suggesting that the soft X-rays are
efficiently absorbed likely due to larger absorption column densities.
Contribution to the FUV lines can also arise from the X-ray irradiated
hemisphere of the secondary star and from material located in
different parts of the flow \cite{gaensickeetal98-2}, where plasma
conditions can be very different from each other. An important
improvement can be achieved by obtaining a systematic survey of
phase-resolved (white dwarf spin and orbital period) of the FUV lines
in magnetic CVs to perform Doppler tomogram analyses and to identify
the kinematical properties of the accretion flow as well as to
separate the different contributions of emission lines. This can allow
a proper test of line formation theory as well as to understand the
irradiation effects of the secondary star.

(2) Magnetic systems have the most complex accretion geometry which is
very difficult to parametrise. Spectral energy distributions (SEDs)
from optical through the UV to the X-rays are necessary to determine
the energy budget and to infer the mass accretion rates
\cite{eisenbartetal02-1}. In the case of truncated discs as in the
moderately magnetised systems, the UV SED is crucial to assess
temperature profile and extension of disc down to the magnetospheric
radius as the SED might show a turn-of in the UV range. Also, the
X-rays can be substantially absorbed leading to accretion luminosities
which can be much lower than those determined with combined UV and
optical observations \cite{mukaietal94-1}. Up to date only a handful
of bright magnetic CVs have been observed in the UV so far, but a
systematic UV study has the potential to infer the relation between
mass accretion rate and system parameters such as inclination angle
and magnetic moments.

Furthermore, it is of fundamental importance to separate the spectral
contributions of the heated pole caps of white dwarfs in polars from
the unheated underlying white dwarf, thereby determining both the
effects of irradiation and the white dwarf temperature. In this
respect an important issue is the tendency of white dwarfs in the
magnetic CVs to be cooler than those in non magnetic systems
(\opencite{sion99-1}, \opencite{araujo-betancoretal05-2}) with
significant differences at all orbital periods. This trend might
reflect a difference in the mass transfer rate efficiency with respect
to non-magnetic CVs as the systems evolve.  In particular, the white
dwarf magnetic field may reduce the secondary star magnetic braking
efficiency \cite{wickramasinghe+wu94-1}, a hypothesis which
observations seem to confirm. However, a statistically significant
sample of magnetic CVs is needed to be observed especially at periods
above the orbital 2--3\,hr period gap where only two systems have been
covered so far with \textit{HST}. This range of the period distribution is
essential as it is dominated by angular momentum loss through magnetic
braking.  It is therefore important to determine the temperature of
the unheated white dwarf atmosphere by means of low resolution UV
spectroscopy over a wide wavelength range either when these systems
are in a low accretion state or via phase-resolved observations which
can allow to isolate the heated atmospheric pole from the unheated
white dwarf atmosphere.

\medskip
\textbf{Future prospectives of UV astronomy.}  To map the accretion
flow structure will need systematic phase-resolved UV spectroscopy at
the orbital and white dwarf rotational periods in high and low
dispersion to study the phase dependence of the FUV emission lines and
of the SED. In particular the determination of kinematical properties
of the accretion flow and the identification of the X-ray irradiated
secondary star atmosphere require the coverage of the various FUV
emission lines and hence a minimum range, of 1150--1800\,\AA\
(1000--1800\,\AA \,desirable) with a spectral resolution of
$R\sim20\,000$. Only for a handful of bright magnetic systems high
resolution spectroscopy has been performed with \textit{HST} and
\textit{FUSE}. Furthermore, the construction of SEDs over the widest
spectral range from the Lyman edge down to 3400\,\AA\ is essential to
determine simultaneously the different spectral components (disc,
white dwarf, accretion funnels). This requires low resolution
spectroscopy at $R\sim2000$ and good quality spectra at levels of a
few $10^{-16}$\,\ecsa. Phase--resolved spectroscopy also demands large
throughput in order to achieve reasonable signal-to-noise ratio with
short exposure times.  Timing capabilities of instrumentation
(e.g. photon--counting systems) allowing to explore different types of
variability (periodic, quasi-periodic and non periodic) on a wide
range of timescales are essential. These can allow the access to the
mostly unexplored temporal domain of UV emission in accreting magnetic
systems.

\section{Accretion discs}
In order to form stars and galaxies, or to power active galactic
nuclei and gamma-ray bursts, matter must be compressed by many orders
of magnitude in size. This is possible while gravity dominates over
thermal, magnetic and rotational energy. This can require the
radiation of substantial amounts of thermal energy and the diffusion
of magnetic field, but ultimately rotation always puts a brake upon
this process because in a homologous collapse of a cloud of size $R$
the rotation energy scales as $R^{-2}$ while gravitational energy
scales as $R^{-1}$. Nature's solution to this problem is to
re-distribute the angular momentum in an accretion disc. In an
accretion disc, gas travels in near-circular orbits gaining angular
momentum from material at smaller radii, and losing it to matter at
larger radii. The transport of angular momentum is driven by some form
of viscosity, and only in recent years has a plausible candidate for
this been identified in the magneto-rotational instability
\cite{balbus+hawley98-1}. Despite this progress, our understanding
of the viscosity of accretion discs and precisely how energy is
dissipated within them remain the central unanswered questions in the
field. A major obstacle to making progress is that two important
properties of discs, their luminosity and temperature distribution,
are independent of viscosity in steady-state discs. Progress can only
be made through the study of phenomena that change on the
\textit{viscous timescale}, $t_\nu \sim R^2/\nu$ where $R$ is the size
of a disc and $\nu$ is the kinematic viscosity or by examining the
vertical temperature structure of discs through their spectra. The
great advantage of close binary stars is their small scales which lead
to viscous timescales of only a few days or weeks, making them
amenable to direct observation.

The outbursts of dwarf novae are almost universally believed to be
driven by changes in the viscosity of the material in their accretion
discs. In the standard \textit{disc instability model} developed in the
1980s, in the quiescent state, the viscosity $\nu$ is very low, and
the viscous timescale, $t_\nu$, is so long that the disc cannot cope
with the rate at which matter flows in at its outer edge. Instead,
mass piles up in the outer parts of the disc until a critical point is
reached and at some radius in the disc the viscosity (and therefore
viscous dissipation rate) increases dramatically, by of order 100
times. This jump can take only a few minutes, with the outburst
following as a heating wave propagates to all radii within the disc. A
major goal of the study of accretion discs is to understand these
outbursts: how they propagate, what triggers them, but above all, why
the viscosity ramps up so violently. This is thought to be rooted in
the ionisation of hydrogen (or in 'ultra-compact' binary stars,
helium), but we have no detailed physical mechanism from which we can
compute the viscosity for a given composition, density and
temperature. All current models, which have been applied to accretion
discs in a wide variety of objects, are hence purely phenomenological.

\begin{figure}
\centerline{\includegraphics[angle=270,width=0.95\textwidth]{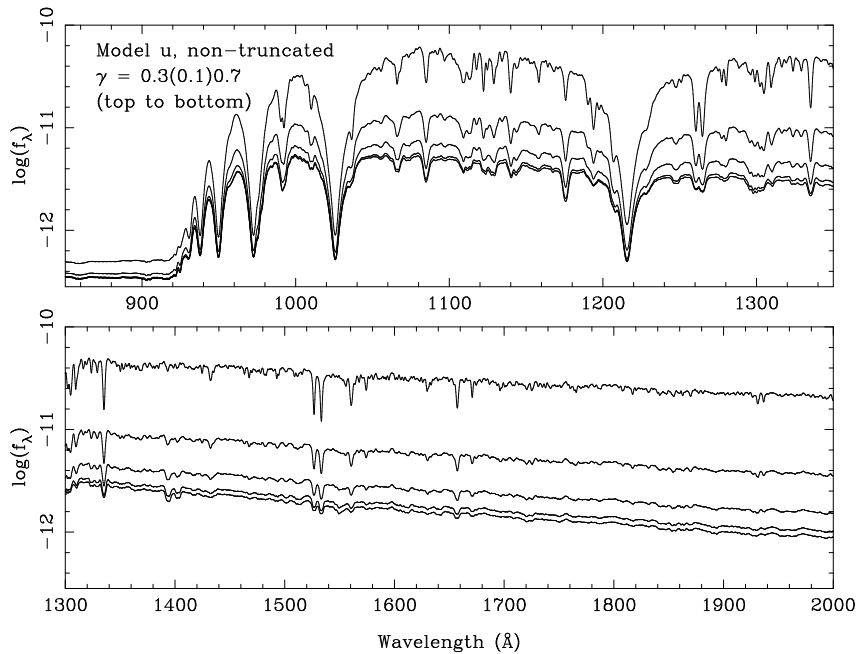}}
\caption{Model spectra of accretion discs with a range of radial
  temperature profiles, $T_\mathrm{eff} \propto R^{-\gamma}$
  \protect\cite{orosz+wade03-1}.}
\label{fig:spectra}
\end{figure}

The commonest, nearest and most easily studied accretion discs are
those of the cataclysmic variable stars which have white dwarf
accretors. Accretion discs around white dwarfs vary in temperature
from $\sim 6\,000$\,K in their outermost parts to over
$100\,000$\,K close to the white dwarf. The radius of the outer disc
is typically 10 to 50 times that of the white dwarf, but it is from
the hot, inner few white dwarf radii that most of the energy is
released. The UV is the key waveband for seeing these
regions. There are three other reasons for UV observation of
accretion discs in these binaries. First, one can see absorption lines
from the disc photospheres most easily in the FUV
(Fig~\ref{fig:spectra}).

These give a handle upon ionisation state not available at optical
wavelengths where emission from the (barely understood) disc
chromosphere is always dominant and photospheric absorption lines are
weak. Second, the UV is where disc model atmospheres currently seem to
fail most severely, in general appearing too blue compared to
observations \cite{orosz+wade03-1}.  The final unique feature of the
UV is its sensitivity to the geometry of the disc because absorption
of the inner by the outer disc is most easily seen in the UV and was
first established from UV observations \cite{horneetal94-1}.

\subsection{Modelling the spectra of accretion discs}

Since the early days of black-body models, followed by models based
upon sums over standard stellar atmospheres \cite{wade84-1}, accretion
discs models based upon modern model atmosphere codes have been
developed \cite{wade+hubeny98-1,orosz+wade03-1}. Such models are
however undermined by our ignorance of the mechanism of viscosity and
hence of the vertical temperature structure of discs. In principle,
spectra can be used in reverse to determine vertical structure, but,
so far, little progress has been made in this area. A problem of long
standing is that disc model atmosphere spectra do not work very well,
especially at FUV wavelengths. This might be down to the vertical
structure, or it could be that the steady-state (radial) temperature
distributions used so far are not accurate \cite{orosz+wade03-1}, even
though it is hard to understand how this can be the case in systems
that hardly change on many viscous timescales. The problem
comes from the small size of accretion discs in close binary stars,
with a typical radius of a few $10^{10}$\,cm.  While their sizes help
keep viscous timescales small, so that thermal instabilities are
easily observed over the course of days to months, it means that the
discs are not spatially resolvable as they subtend at best a few
$\simeq0.01$ milli-arcseconds for the closest systems. Direct imaging
of the accretion discs will be possible in the foreseeable future only
in the much larger symbiotic stars (Sect.\,\ref{s-symbiotic}), in
which, however, the viscous time scales are substantially longer, and
the temporal variability of symbiotics in terms of disc instabilities
is much less established compared to the situation in CVs. Thus the
spectra we see are the integrated spectra from all radii in the
disc. This makes it hard to know whether it is the radial or vertical
structure that is causing the problem. With integrated spectra, one
cannot be sure whether all radii are poorly modelled or whether only
specific effective temperatures are involved, and thus it is not clear
how to adapt models.  We need spatially resolved spectra, which can be
achieved through a technique known as eclipse mapping
\cite{horne85-1}. Applied at UV wavelengths, this technique has the
capability to provide spatially resolved spectra of the inner parts of
accretion discs where most energy is released. We can then see where
it is in the discs that model atmospheres fail most severely.

\begin{figure}
\centerline{\includegraphics[width=0.75\textwidth]{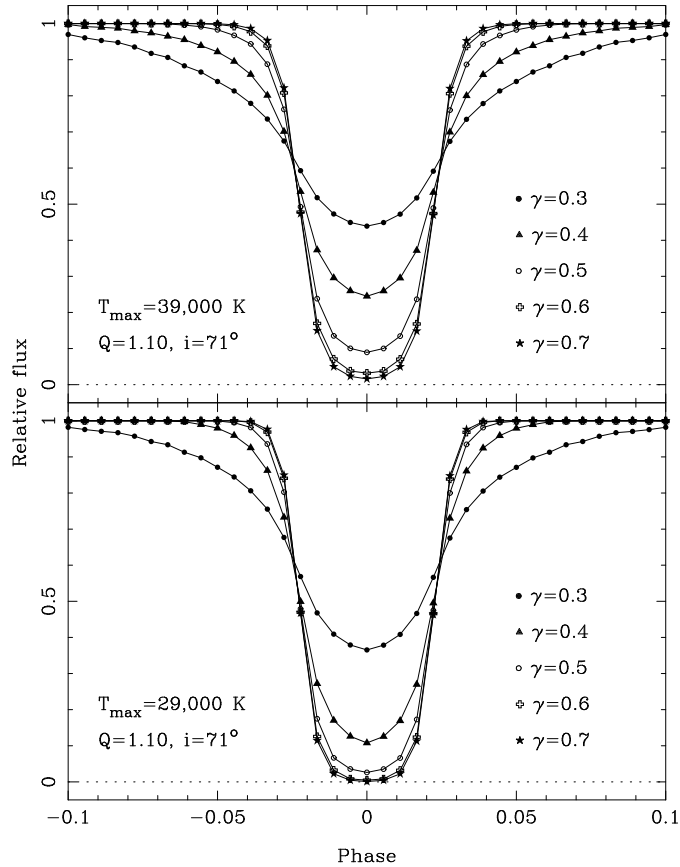}}
\caption{Model eclipses in the continuum from 1410 to $1530\,$\AA\ for
black-body (top) and model atmosphere discs for a range of radial
temperature profiles, $T_\mathrm{eff} \propto R^{-\gamma}$
\protect\cite{orosz+wade03-1}.}
\label{fig:eclipses}
\end{figure}

The principle of eclipse mapping is as follows: in an eclipsing
system, the light-curve of the disc as it is eclipsed depends upon how
concentrated the surface brightness is. For instance a flat
distribution of brightness leads to a shallow V-shaped light curve,
whereas a distribution which is strongly peaked towards the centre of
the disc has a deep U-shaped light curve (Fig.~\ref{fig:eclipses})

\begin{figure}
\centerline{\includegraphics[width=0.75\textwidth]{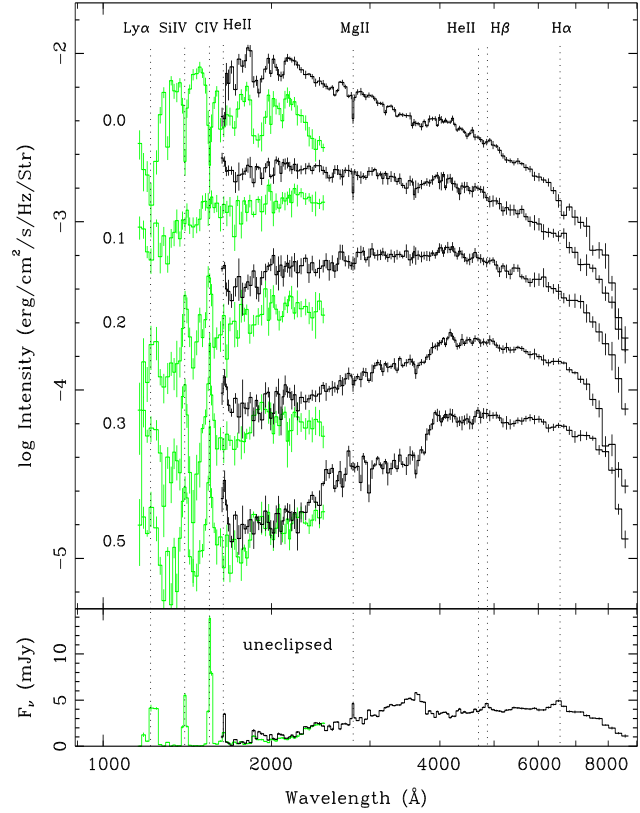}}
\caption{The spectra of the steady-state system UX\,UMa as a function
of radius deduced from \textit{HST}/FOS observations of its eclipse
\protect\cite{baptistaetal98-1}. The spectra are those of annuli with
central radius indicated in units of the distance to the inner
Lagrangian point. Note that the spectra are plotted in $f_\nu$; the UV
is dominant energetically.}
\label{fig:baptista}
\end{figure}

Thus, in essence, the light curves can be used to deduce the variation
of surface brightness with radius. Eclipse mapping was developed by
\inlinecite{horne85-1} for broad-band optical light curves. A significant
step in eclipse mapping came with its extension to spectra
\cite{ruttenetal93-1,ruttenetal94-1}. The simple, beautiful, idea was
to carry out eclipse mapping on each pixel of low-resolution spectra
to produce spectra at every point of discs. This technique applied to
UV data has the capability of giving us spatially-resolved spectra of
the inner accretion discs of close binary stars. Only one such
analysis has been carried out in the UV with \textit{HST}/FOS
(\opencite{baptistaetal98-1}, Fig.~\ref{fig:baptista}) of the brightest of
all high-state systems, UX\,UMa.

Even on this, the brightest suitable system, the study was limited by
signal-to-noise ratio, especially at FUV wavelengths, precisely the most
important part of the spectrum. The signal-to-noise ratio in this region of
the spectrum is a modest 10\% and yet the radial resolution is still
only $\sim 4$ white dwarf radii, which means that we are still
seeing the integrated light from a region which varies by a factor of
three in temperature from the inner to outer edge of the annulus. In
other words the problem of integrated spectra is only partially solved
in this study. 

\medskip
\textbf{Future prospectives of UV astronomy.} 
To substantially improve our ability to model the spectra of accretion
discs  requires a low resolution, wide
wavelength coverage UV spectrograph of much greater sensitivity than
has been available to date. Low resolution ($R \sim 300$) because the
spectral eclipse mapping technique cannot resolve the $\sim
1000\,\kms$ motions within the disc. UV because, as said before, this
is where most luminosity is radiated and where the current disc
atmospheres fail most severely. Wide wavelength coverage (at least
1000 to 3000\,\AA, and if possible extending to the Lyman edge) because it is
the variation of continuum flux with wavelength which tells us most
directly about the vertical structure in stellar atmospheres. Finally,
and perhaps above all, in comparison to any UV mission to date, high
sensitivity is needed in order to improve both the signal-to-noise ratio in
the deconvolved spectra and their radial resolution down to of order a
single white dwarf radius and so that the method can be applied to
systems fainter than UX\,UMa. Two other requirements needed for this
work are an ability to take short exposures ($<2$ seconds) which are
accurately timed with absolute times good to better than one-hundredth 
of the exposure length.

Once progress in understanding gross properties of spectra has been
made, there will be a need for higher resolution observations. Models
can predict the changes in detailed line profiles expected during
eclipse \cite{orosz+wade03-1}. Again disc broadening means that moderate
resolution is sufficient ($R \sim 5000$), but high-sensitivity in
order to allow short exposures and therefore high spatial definition
of the disc are a must. The study of integrated spectra provides the
most stringent requirement for spectral resolution because face-on
discs (no eclipse) have significantly narrower line profiles and
suffer less from blending. For these $R\sim20\,000$ would be
useful, covering from 900\,\AA\ to 1700\,\AA.

\subsection{Disc instabilities}
\begin{figure}
\centerline{\includegraphics[width=0.95\textwidth,angle=270]{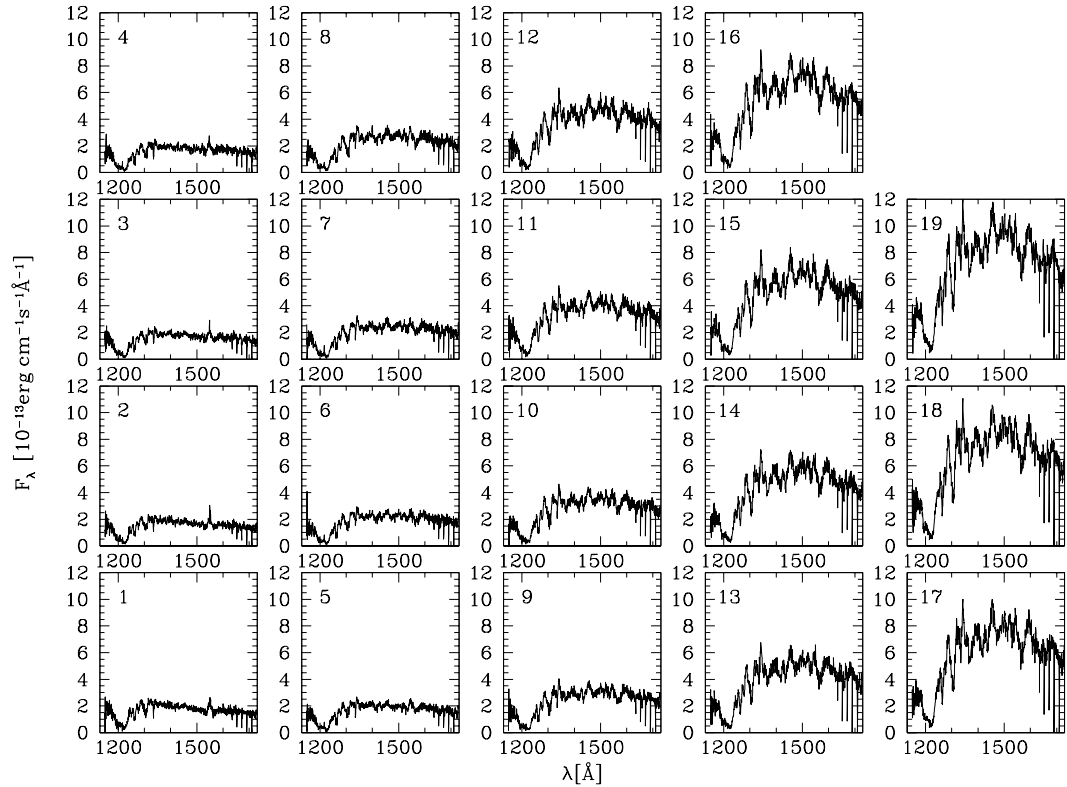}}
\caption{The spectra of the dwarf nova VW~Hyi caught at the start of
  an outburst with \textit{HST/STIS} \protect\cite{sionetal04-2}. At first
  the spectra are dominated by the white dwarf but strong photospheric
  absorption lines develop as the heating front reaches the inner
  disc.}
\label{fig:vwhyi}
\end{figure}

It is the study of dwarf nova outbursts that lead to the disc
instability theory and the discovery of the strong dependence of
viscosity with the physical conditions within the disc. The disc
instability model has largely been used to explain the gross features
of outbursts, such as their duration and amplitude, but there have not
been convincing detections of the heating fronts which would allow us
to confirm predictions of the models in detail. Attempts have been
made from optical observations of eclipses, but these lose resolution
in the inner disc because the outer disc dominates the light output at
optical wavelengths. The propagation of the heating fronts into the
inner disc is of particular interest because there is evidence to
suggest that the inner disc is strongly depleted during quiescence
\cite{schreiberetal04-1}. This is needed to prevent outbursts
triggering in the inner disc, which in some systems would lead to too
high an outburst rate.  Propagation of heating fronts can be measured
from the development of photospheric line profiles from the disc. As
the front progresses inwards, the contribution from smaller radii in
the disc will contribute to broadening the line profiles because
Doppler broadening is largest in the inner disc. The photospheric
lines are strongest by far in the UV and develop dramatically during
outburst (Fig.~\ref{fig:vwhyi}).

\medskip
\textbf{Future prospectives of UV astronomy.} So far this sort of
study has not been possible because of limited sensitivity at FUV
wavelengths.  It also requires a much higher duty cycle than possible
with \textit{HST} as outbursts take of order a few hours to a day to
start. Observations such as these could also show the development of
winds through the resonance lines which are only visible at UV
wavelengths. A spectrograph covering 900 to 1700\,\AA\ with a resolution
$R > 5000$ is needed for this work.

\section{Hydrogen-deficient systems}
A unique aspect of close binary stars is that owing to the evolution
of their mass donor stars, some systems can show very unusual
abundances, adding an extra dimension to the development of
atmospheric models. The AM\,CVn systems for example have helium white
dwarf donors and accretion discs which are $>95$\%
helium. Ultra-compact neutron star/white dwarf binaries can have
carbon-oxygen and oxygen-neon-magnesium donor stars. The element
abundances are crucial to understanding the evolution that leads to
such stars. For instance, an evolutionary path from cataclysmic
variable stars to AM\,CVn stars typically leaves a small amount of
hydrogen \cite{podsiadlowskietal03-1}, whereas a route via double white
dwarf mergers does not. Similarly, the ratios of the CNO elements in
such stars depends upon the initial mass of the donor stars. Such
information is invaluable in pinning down evolutionary pathways, and
therefore to predicting the numbers of systems. These binaries are so
compact that they can fit comfortably inside the Sun. At the same time
their short orbital periods means that gravitational radiation is
strong (such systems will be significant sources for \textit{LISA}) and
mass transfer rates can be high. As a result, they emit mostly at UV
wavelengths and the UV is where photospheric lines from the disc are
strongest. 

\medskip
\textbf{Future prospectives of UV astronomy.}  Over the next few years
many more AM\,CVn systems are likely to be discovered, but as they are
relatively rare, the majority will be faint. As for the hydrogen-rich
systems, a $R \sim 20\,000$ spectrograph covering 900\AA\ to 1700\AA\
is needed, with high sensitivity the key feature.

\section{Accretion winds}
Mass loss is an ubiquitous feature of astrophysical systems and the
evidence of mass loss in disc-dominated cataclysmic variables is
unambiguous.  In the wavelength range accessible to \textit{IUE}
($\simeq1150-3200$\,\AA), the existence of outflows in systems
observed a lower inclination is indicated by P-Cygni-like and/or blue
shifted absorption profiles in resonance transitions of \Ion{N}{V},
\Ion{Si}{IV}, and most commonly \Ion{C}{IV}.  Velocity widths of
3000--5000\,\kms, comparable to the escape velocity from the primary,
are regularly seen, especially in \Ion{C}{IV}.  The features are
understood to result largely from scattering of disc photons by the
outflow.  At low inclinations, the process removes photons along the
line of sight to the disc; emission wings arise from photons scattered
into the line of sight of the observer, just as in the stellar winds
of massive stars.  At higher inclinations, less direct light is
observed from the disc, and the resonance lines generally appear as
broad emission features.  Indeed, after analysing 850--1850\,\AA\
spectra of Z\,Cam obtained with the \textit{Hopkins Ultraviolet
Telescope} \inlinecite{kniggeetal97-1} suggested that virtually all of
the lines in the UV spectrum of a typical high-state CV are formed in
the outflow, either in the supersonic portion of the wind or in a
lower velocity portion of the wind near the interface with the disc
photosphere.

Although the strong 1s--2p transitions of Li-like or Na-like ions
dominate the line spectra of disc-CVs observed with \textit{IUE} and
\textit{HST}, the FUV spectra obtained with \textit{FUSE} often show narrower
features from intermediate ionisation state transitions of abundant
ions such as \Ion{N}{III}, \Ion{C}{III}, \Ion{Si}{III}, and
\Ion{Si}{IV}.  These intermediate level ionisation state lines often
show orbital phase dependent effects, even in systems of intermediate
inclination such as Z\,Cam \cite{hartleyetal05-1} that suggest the
effects of the accretion stream must be included to complete the
picture of extra-planar gas in disc-dominated systems.  In RW\,Sex and
V592\,Cas, enigmatic orbital variations in the blue edges of  broad
\Ion{C}{III} profiles indicate departures from bi-conical symmetry in
the high-velocity wind \cite{prinjaetal03-1,prinjaetal04-1}. Whether
these are associated with disc tilts or the accretion stream or some
other mechanism is not understood. Unfortunately the number of systems
in which appropriate studies have been undertaken is small, and
generally speaking not intensive or lengthy enough to fully
characterise the phenomenology of the effects.

\begin{figure}
\centerline{\includegraphics[angle=270,width=0.95\textwidth]{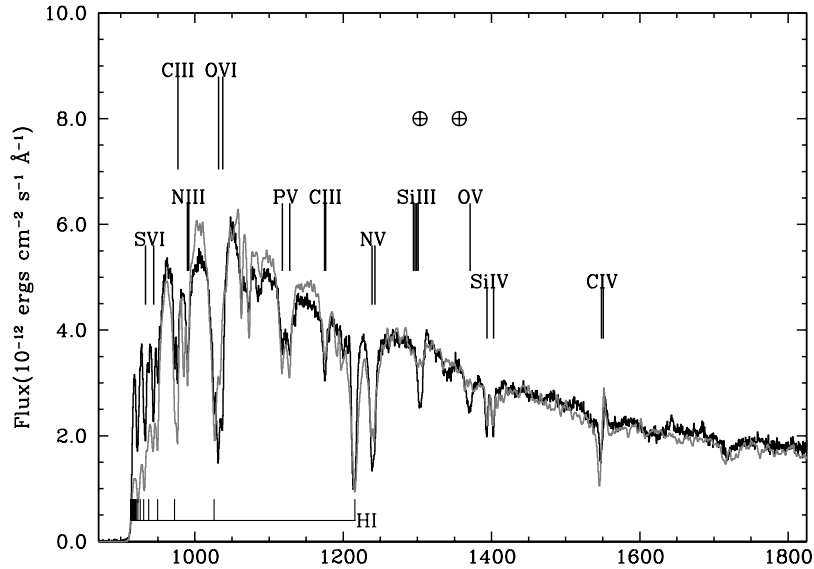}}
\caption{\textit{HUT} spectrum of the wind dominated spectrum of IX\,Vel
modelled compared to one of the models of \protect\inlinecite{long+knigge02-1}}
\label{f-long}
\end{figure}

Originally, the possibility that the wind was a radial wind was
considered, but observations of eclipsing systems showed changes in
profiles shapes that are most straightforwardly interpreted as an
indication of rotation, thereby indicating that the wind emanates from
the inner disc \cite{drew87-1}. Consequently, our basic picture of the
high velocity wind first observed with \textit{IUE} is of a bi-conical
flow emanating from the inner portion of the disc and/or rapidly
rotating boundary layer.  \inlinecite{vitello+shlosman93-1} were the
first to attempt to actually model the profile shapes of wind lines as
observed in high state CVs in terms of kinematic prescription for a
bi-conical wind. They found that the \textit{IUE}-derived ($R=200$)
\Ion{C}{IV} profiles of three systems~--~RW\,Sex, RW\,Tri, and
V\,Sge~--~could be reproduced with moderately collimated winds with
the local mass loss rates of order 10\% of the disc accretion rate and
terminal velocities of 1--3 times the escape velocity at the footprint
of each streamline.  Subsequently, \inlinecite{knigge+drew97-1} succeeded,
using a somewhat different kinematic parameterisation for a bi-conical
flow, in reproducing the \Ion{C}{IV} profile of UX\,UMa through an
eclipse. This analysis was important, not only because it was the
first attempt to model changes in the profile through eclipse, but
also because it suggested, at least in UX\,UMa, the existence of
a relatively dense, high column density, slowly outflowing transition
region between the disc photosphere and the fast moving wind.  Both
the Vitello \& Shlosman and Knigge \& Drew analyses suggested that the
characteristic acceleration length for the high velocity winds
observed in disc dominated CVs is quite long, or order
$100\,R_\mathrm{wd}$. Most of the analyses of the spectra of disc winds
were limited to single lines, but more recently \inlinecite{long+knigge02-1}
have developed Monte Carlo radiative transfer codes which in a few
cases (see Fig.\,\ref{f-long}) are able to qualitative reproduce the
full UV spectrum of a disc dominated CV.  Hydrodynamical simulations
of radiatively-driven CV winds are also been undertaken, and when
combined with a radiative transfer code, these are also beginning to
be compared to observed spectra with mixed results (see, e.g.,
\opencite{proga03-1}).

Although, modelling of CV winds has progressed, fundamental basic
questions about the winds still remain. We are unable to measure basic
parameters like the mass-loss rate, and although the wind is assumed
to be radiatively driven, the observational and theoretical evidence
for this is at best murky.  For example, on the observational side, if
the wind is radiatively driven, one  might expect that the
observational signatures of wind lines would be strongest when systems
are brightest.  But \inlinecite{hartleyetal02-1} found there was no
correlation between the strength of wind features and continuum
brightness in the spectra of three observations each of the two
nova-like variables IX\,Vel and V3885\,Sgr with \textit{HST}.  (Unfortunately, the
number of high state systems that have been observed enough to begin
to characterise their behaviour with time is quite limited.) And
\inlinecite{drew+proga00-1} have argued that the luminosity of discs
is at best marginally enough to accelerate a high velocity wind.
Thus, alternatives to the emission or additions to radiation pressure
must be considered instead. These include viscous heating of the upper
portion of the disc atmosphere \cite{czerny+king89-1}, and irradiation
\cite{czerny+king89-2} as well as magneto-centrifugal forces producing
constant angular velocity out to the Alven surface
\cite{cannizzo+pudritz88-1}.

\medskip
\textbf{Future prospectives of UV astronomy.}
As observations of high mass transfer discs and winds in CVs have improved, so
has the complexity of phenomenological descriptions of the wind
structures emanating from the disc, especially as the wavelength
coverage has extended in the the UV.  But very few systems have
been studied in enough detail to isolate common from uncommon
behaviour.  Furthermore, it is quite clear that the appearance of
disc dominated systems is strongly modified by inclination, and as
a result one needs to observe a number of similar systems to the
same level of detail to be able to go beyond the general
variations that were observed with \textit{IUE}.  To carry out a study of
this type  higher sensitivity is required so that the pool of
targets that can be studied is substantial and so that the
observations can be made at the resolution needed to resolve the
narrower lines that exist particularly in the FUV short-ward of
1200\,\AA.

Higher sensitivity observations are also required to obtain a better
short term characterisation of the wind flow.  Some systems seem
to have little or no short term temporal variability, whereas
others, e.g BZ\,Cam \cite{prinjaetal00-1} are highly variable. We do
not know whether this is due to some fundamental difference in the
systems~--~a magnetic white dwarf for example~--~or is it due simply to
differences in the accretion rate.  What is the role of outer disc
in the creation of disc wind? Some hydrodynamical simulations show
fast steady flows emanating from the inner disc, but complex time
variable flows in the outer discs.

More systems need to be measured with high signal-to-noise ratio.  The
number of systems actually observed with \textit{HST} were far fewer
than observed with \textit{IUE}, a fact that was partially a result of
the way \textit{IUE} was scheduled compared to \textit{HST} and the
fact that \textit{HST} was never designed to be a dedicated UV
observatory, but a multi-purpose / multi-wavelength facility.  Instead
the observations with \textit{HST} have focused on a few key systems.
Therefore it has been quite difficult to determine whether many of the
phenomenological models proposed to explain the wind features of disc
dominated features are founded on general characteristics of winds in
disc dominated CVs and how many are due to individual systems.

To maximally constrain models of the wind, the wavelength coverage of
a new mission should extend to the region containing \Ion{O}{VI}, and
possible to the Lyman limit.  Including \Ion{O}{VI} (along with
\Ion{N}{IV}, \Ion{Si}{IV}, \Ion{C}{IV}, and \Ion{He}{II}) is important
not only because \Ion{O}{VI} represents the next step up in the
temperature space ladder, but also because the FUV below 1150\,\AA\
is rich in lines of intermediate ionisation states.  These
lines establish stringent constraints on physical conditions in the
region near the disc plane.

\section{Black-hole binary stars}

Most of the dynamically-confirmed black hole binaries are transients,
which spend most of their time in a low-luminosity state.  Recently
there has been much debate surrounding comparison of these quiescent
black holes with their neutron star analogues in the attempt to detect
"direct" evidence of event horizons in the former systems. Neutron
stars are brighter in X-rays, as might be expected if the black holes
advect accretion energy through the event horizon.  The theoretical
models for low-luminosity accretion flows onto black holes, however,
include variants where the flow is unbound so that much of the
accretion energy may be carried away as kinetic energy of an outflow.
Hence it is crucial to identify the correct theoretical model before
claiming event horizons have been detected.  The UV is a vital window
for achieving this: almost any model can reproduce the X-ray data
alone by varying the fit parameters, but simultaneously fitting the UV
spectra is much more exacting while optical and infrared wavelengths
are hopelessly contaminated by the donor star and outer disc. The
problem is that these systems are faint in quiescence and only three
quiescent UV spectra exist to date: of the black holes A\,0620-00 and
XTE\,J1118+480 and of the neutron star Cen\,X-4.  The black hole spectra
resemble each other and differ markedly from Cen\,X-4's.  This suggests
a real physical difference, but clearly insufficient to decide
definitively between models. We need to observe more systems, and
obtain simultaneous X-ray and UV data.

In outburst, transients brighten by factors of $\sim 1000$ across the
optical-UV-Xray spectra regions.  This is attributed to the same disc
instability that drives the outbursts of cataclysmic variable stars,
but the black-hole binaries are complicated by (i) irradiation of the
optical-UV emitting disc by the central X-ray source which changes the
effective temperature distribution and causes warping of the disc,
(ii) by large discs which have no global stable high-state
configuration below the Eddington limit, (iii) for reasons which are
not yet fully understood, the inner accretion disc is often missing,
being replaced within a transition radius, $R_\mathrm{tr}$, by an
optically thin, inefficiently-radiating advective flow.  These factors
substantially alter the character of the sources: luminosity
generation, outflows, duty cycle, and the mass accumulation by the
black hole are all changed.  To understand these complications, UV
observations are essential: the optical is dominated by the outermost
disc (which behaves more like CV discs) and by the donor stars. The UV
is required to see unambiguously the signatures of irradiation in the
SED, to detect self-occultation by warping, and to measure the
transition radii.

\medskip
\textbf{Future prospectives of UV astronomy.} Only a handful of black
hole X-ray transient outbursts have had their SEDs monitored
throughout the outburst, and their behaviour has been
diverse. Transients outburst on time-scales of decades and there are
many that we have yet to detect. To understand the outbursts in a
systematic way, more SED monitoring, including the UV, is required. 
Broad wavelength coverage at low resolution and high throughput are
essential for this kind of studies.

\section{\label{s-globulars} Star clusters as laboratories for close
binary evolution} 

It has been known since the mid-1970s that there is a 100-fold or so
overabundance of bright LMXBs in globular clusters (GCs), relative to
the galactic field (e.g. \opencite{katz75-1}). This quickly led to the
realization that the high stellar densities in the cores of GCs might
open up entirely new {\em dynamical} channels for the formation of
interacting close binaries. The most famous of these is tidal
capture, a 2-body process resulting from a close encounter between a 
compact object (white dwarf or neutron star) and an ``ordinary''
cluster members (main sequence star or giant). During such an
encounter, the latter star experiences tidal distortions. This 
dissipates orbital energy and can therefore lead to capture and binary
formation \cite{fabianetal75-1}. However, interacting
binaries can also be formed via 3- and 4-body interactions,
i.e. processes involving existing binaries. For example, in a close
encounter between a low-mass (e.g. MS/MS) binary system and a high
mass (e.g. NS) single star, the most likely outcome is ejection of the
lowest mass participant and formation of a NS/MS binary system
\cite{sigurdsson+phinney93-1}. 

Interacting binaries in GCs deserve careful study for two basic
reasons. First, they can in principle provide us with large,
uniformly-selected samples of systems at known distances. This is
precisely what is needed to test theoretical binary evolution
scenarios. Second, close binaries are actually key players in
controlling the late dynamical evolution of GC themselves. Thus
interacting binaries can actually be used as tracers of the
dynamically-formed close binary population in observational studies of
GC evolution. In practice, the inevitable feedback between binary and
cluster evolution will complicate things, but there is no doubt that
interacting binaries in GCs can provide us with unique insights into
both types of evolution. 

\begin{figure}
\centerline{\includegraphics[angle=-90,width=0.8\textwidth]{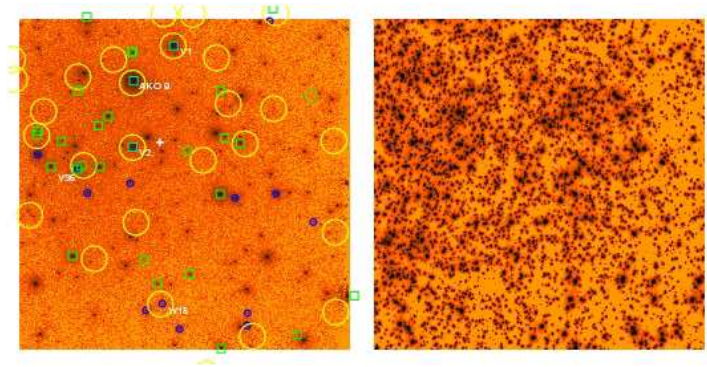}}
\caption{\label{f-gc}{\em Left Panel:} A deep \textit{HST}/STIS FUV
image of the core of 47\,Tuc. The image is approximately
25''$\times$25'' in size and includes the cluster centre (marked as a
white cross). For comparison, 47\,Tuc's core radius is 23''. The
positions of previously known blue objects (green squares), Chandra
X-ray sources (large yellow circles) and CV candidates (small blue
circles) are marked. The four confirmed CVs within the field of view are
labelled with their most common designations. The image is displayed
on a logarithmic intensity scale and with limited dynamic range so as
to bring out some of the fainter FUV sources. {\em Right Panel:} The
co-added \textit{HST}/WFPC2/F336W (roughly U-band) image of the same
field. This image, too, is shown with a logarithmic intensity scale
and limited dynamic range. Figure reproduced from
\protect\inlinecite{kniggeetal02-1} (\protect\copyright~2002 The
American Astronomical Society.)}
\label{f-spectra}
\end{figure}

UV astronomy has a key role to play in this area. Accreting binaries
tend to have much bluer spectral energy distributions than the
late-type main sequence stars that make up the bulk of stellar
clusters and galaxies. This immediately implies that FUV observations
should be an excellent way to find and study these populations, even
in optically crowded fields, such as GC cores. This expectation is
strikingly confirmed in Figure\,\ref{f-gc}, which shows FUV and U-band images
of the same central regions of the GC 47\,Tuc. The difference in
crowding is obvious, and several CVs and new CV candidates pop up
nicely in the FUV image. This image represents the deepest FUV survey
of any GC carried out to date, and utilises observations obtained with
STIS onboard \textit{HST} \cite{kniggeetal02-1}. Earlier generations of FUV/NUV
detectors on \textit{HST} have also been used to search for and study
interacting binaries in GCs 
(e.g. \opencite{paresceetal92-1}, \opencite{demarchietal93-1},
\opencite{ferraro+paresce93-1}, \opencite{demarchi+paresce94-1},
\opencite{demarchi+paresce96-1}, \opencite{paresce+demarchi94-1},
\opencite{cooletal95-1}, \opencite{sosin+cool95-1}).
In the case of 47\,Tuc, the lack of crowding in
the FUV even makes it possible to carry out slitless, multi-object
spectroscopy in the cluster core \cite{kniggeetal03-1,knigge04-1}.

In principle, open clusters and local group galaxies could also be used
as binary evolution laboratories. However, open clusters contain fewer
stars than GCs and are characterised by lower central densities. Thus
interacting binaries are not as abundant in open clusters as in GCs,
and the construction of a statistically interesting sample would
probably have to involve studies of many such clusters. Local group
galaxies obviously harbour large interacting binary population as
well. However, even with a 4\,m class space telescope, UV observations
reaching the depths required to study the quiescent interacting binary
populations will be extremely challenging for the Magellanic Clouds
and probably impossible for all other local group galaxies.

\medskip
\textbf{Future prospectives of UV astronomy.} Several additional
galactic GCs have recently been imaged in the UV with \textit{HST}, so the UV
picture of their interacting binary populations will become clearer as
soon as these new data sets have been analysed. However, all of these
studies are seriously constrained by the small field of view of both
the STIS and ACS UV detectors (roughly 30''$\times$30''); this often
makes it impossible to obtain a complete census of the interacting
binary population. For example, the deep UV image of 47 Tuc in
Figure\,\ref{f-gc} covers only about 1/3 of the cluster core. GALEX will be of
some use in this regard (e.g. to find sources in GC outskirts and open
clusters), although the benefit of its larger field of view is
partially offset by its poorer spatial resolution and lower
sensitivity (relative to \textit{HST}).

The optimal future UV imaging instrument would consist of a large
($\ge4$\,m) mirror feeding a large-format detector producing images
with diffraction-limited spatial resolution. However, the ability to
obtain spectral information is also crucial to allow secure
classifications of the detected UV sources. Single-slit/single-object
spectroscopy is an extremely inefficient way of obtaining this
information in a cluster setting. As noted above, slitless
spectroscopy may be used in special cases, but what is really needed
is a more generally applicable way to carry out multi-object
spectroscopy (MOS) in the UV. MOS using optical fibres is probably not
an option, since fibre losses rise steeply towards short wavelengths
(at least in the current generation of fibres). Configurable slit
masks are probably also impractical in a space-based observatory,
since their use would require a large number of delicate moving
parts. A simple, low-tech solution is to provide a reasonably large
selection of narrow-band filters. An intriguing high-tech solution
might involve superconducting tunnel junction (STJ) detectors
(e.g. \opencite{cropperetal03-1,verhoeve02-1}, see also
\opencite{romanietal99-1}). These are able to provide an energy
estimate for every photon detected, so imaging and spectroscopy could,
in principle, be done in a single observation.

\section{Requirements on future UV instrumentation}

Here we summarise the instrumental requirements defined by the
scientific goals above.

\begin{enumerate}
\item Low-resolution spectroscopy ($R\simeq1000-2000$), with a
wavelength coverage as large as possible. Optimum would be
simultaneous data from Lyman edge down into the blue optical
($\simeq5500$\,\AA). The first priority is the broadband coverage, and
highest throughput. Continuum signal-to-noise ratio of $\simeq10$ at
flux levels of a few $10^{-16}$\,\ecsa\ should be achieved short
exposures ($10-30$\,min).
\item Medium-resolution spectroscopy ($R\sim20\,000$). The ``standard''
wavelength range $1150-1800$\,\AA\  would be adequate, covering the
entire far UV down to the Lyman limit would be preferable. 
\item Detectors. Both low and medium spectrographs should have photon
  counting detectors with absolute times accurate down to fractions of
  a second. 
\item Large field-of-view UV imager (10\,arcmin) with high spatial
  resolution (diffraction limited). Broad-band UV filters. Photon
  counting with accurate timing information. UV/optical
  interferometrie providing sub-milliarcsec spatial resolution.

\end{enumerate}


\end{article}
\end{document}